\documentclass[10pt,conference,letterpaper]{IEEEtran}
\IEEEoverridecommandlockouts

\usepackage{cite}
\usepackage{hyperref}
\usepackage{amsmath,amssymb,amsfonts}
\usepackage{algorithm}
\usepackage{algorithmic}
\usepackage{graphicx}
\usepackage{textcomp}
\usepackage{booktabs}
\usepackage{multirow}
\usepackage{makecell}
\usepackage{pifont}
\usepackage{url}
\usepackage{kotex}
\usepackage{array}
\usepackage{booktabs}
\usepackage{xcolor}
\usepackage{colortbl}
\usepackage{microtype}
\usepackage{subcaption}
\usepackage[normalem]{ulem}
\usepackage{enumitem}
\usepackage{eso-pic}

\def\BibTeX{{\rm B\kern-.05em{\sc i\kern-.025em b}\kern-.08em
T\kern-.1667em\lower.7ex\hbox{E}\kern-.125emX}}

\usepackage{xspace}
\newcommand{\sys}{\textsc{MoSim}\xspace}

\begin{document}
\bstctlcite{IEEEexample:BSTcontrol}

\title{Accurate Simulation of Distributed Training Jobs with Network Contention Modeling}

\author{%
  \IEEEauthorblockN{
    Yeonho Yoo\IEEEauthorrefmark{1},
    Hyunho Lee\IEEEauthorrefmark{2},
    Hyunmok Choi\IEEEauthorrefmark{2},
    Chuck Yoo\IEEEauthorrefmark{2},
    Gyeongsik Yang\IEEEauthorrefmark{2}
  }
  \IEEEauthorblockA{
      \IEEEauthorrefmark{1}Department of Computer Science and Artificial Intelligence, Dongguk University\\
    \IEEEauthorrefmark{2}Department of Computer Science and Engineering, Korea University 
  }
}

\AddToShipoutPicture* {
\AtPageLowerLeft {
\put(0, 40){
\makebox[\paperwidth][c]{
\begin{minipage}{\textwidth}
\centering
\footnotesize
\copyright~2026 IEEE. Personal use of this material is permitted. Permission from IEEE must be obtained for all other uses, in any current or future media, including reprinting/republishing this material for advertising or promotional purposes, creating new collective works, for resale or redistribution to servers or lists, or reuse of any copyrighted component of this work in other works. \\
This paper has been accepted for publication in IEEE MASCOTS 2026.
\end{minipage}
}
}
}
}

\maketitle

\begin{abstract}
Trace-driven simulation is widely used to evaluate distributed training (DT) jobs in GPU clusters, but existing simulators either ignore network contention or approximate it with a fixed penalty. This misses how scheduling decisions determine which jobs share server network interfaces and inter-server links, thereby changing networking time during training. As a result, our motivating experiments demonstrate that they incur large errors, reaching up to 73.64\% mean absolute percentage error (MAPE) in average job completion time (JCT). This paper introduces \sys{}, a GPU-cluster simulator that models DT job execution under dynamic network contention. \sys{} combines GPU-free characterization with network contention model: it obtains each job's compute time, networking time, and networking volume without GPUs, then uses the current worker assignment to estimate how shared network interfaces affect each job's iteration time. Our evaluation shows that, compared with existing simulators, \sys{} reduces simulation error for average JCT by up to 3.28$\times$, tail (99th-percentile) JCT by up to 7.79$\times$, and makespan by up to 8.48$\times$, while modeling NIC contention factors with only 8.63\% error on average. By avoiding real-GPU profiling, \sys{} also reduces input construction overhead by 44.6$\times$.

\end{abstract}


\section{Introduction}
Simulation has become an essential tool for analyzing the execution of distributed training (DT) jobs and the efficiency of GPU-cluster infrastructure, since large-scale what-if studies on real hardware are costly and disruptive. A useful simulator must capture how the training time of active jobs changes under shared cluster resources. This is challenging because DT jobs repeatedly synchronize gradients over the network, and concurrent jobs compete with one another when their traffic shares the same server NICs or inter-server links. As job scheduling determines which jobs run concurrently and which network resources they share \cite{go2026making}, accurately modeling job contention on the network is key to understanding job execution times, cluster efficiency, and resource management decisions in GPU clusters.

In a DT job, gradients are synchronized over the network every iteration, and since gradient size scales with model parameter size, networking alone can occupy up to 90\% of training time \cite{pipedream}. When concurrent jobs are co-located on the same server or routed through the same inter-server link, their synchronization traffic contends for shared bandwidth, and this contention (not additional computation) is what slows each job down (\S\ref{sec:2.2}).
Crucially, this contention is the common case in production, not an artificial corner case. Over 86\% of jobs in the Microsoft Philly cluster \cite{philly} and 93\% of jobs in AcmeTrace \cite{acme} request 8 or fewer GPUs. These small and medium DT jobs are exactly where a scheduler has many choices---keeping a job within a server, spreading it across servers, or running multiple jobs so that their inter-server traffic overlaps---so whether their traffic shares a NIC or an inter-server link is decided at scheduling time, not fixed in advance.

So, accurately simulating DT job performance under such contention is challenging but critical. Yet, in our motivating experiments, existing GPU-cluster simulators incur large errors---73.64\% mean absolute percentage error (MAPE) in job completion time (JCT). This error arises because existing simulators handle contention in two limited ways. No-contention simulators, such as Tiresias \cite{tiresias} and Gavel \cite{gavel}, use isolated job execution times and do not model contention between concurrent jobs. Static-contention simulators, such as Pollux \cite{pollux} and Muri \cite{muri}, approximate contention by applying a fixed factor, e.g., increasing every job's training time by 10\%.
Both are scalable but fail to capture the workload- and scheduling-dependent nature of network contention. 
Our experiment shows that, relative to contention-free execution, network contention increases job training time by up to 114\%, with an average increase of 33\% (\S\ref{sec:motivation}).
Therefore, ignoring it or replacing it with a fixed factor leads to the large errors noted above. This level of error can substantially distort what-if analysis of job execution and cluster efficiency, making the results unreliable for resource-management decisions.

To address the inaccuracy, we introduce \sys{}, a GPU-cluster simulator that estimates DT job execution time under dynamic network contention. Rather than ignoring contention or applying a fixed factor, \sys{} derives contention directly from each job's execution characteristics. For each job, \sys{} constructs the quantities it needs (\S\ref{subsec:input})---compute time, networking time, and networking volume---via \emph{GPU-free characterization}, which obtains them without real-GPU profiling. \sys{} then combines these per-job quantities with scheduling decisions (\S\ref{subsec:des}) and applies \emph{network contention model} that estimates each job's contention on shared NICs and inter-server links (\S\ref{subsec:cluster-level-sim}) and calculates the changing training time. As the cluster state evolves, \sys{} continuously updates each job's networking time, enabling high-fidelity simulation of DT job execution and overall cluster efficiency.
We make the following contributions:
\begin{itemize}
\item Quantification of network contention, showing that training time can increase by up to 114\% (33\% on average) mainly due to networking contention.
\item Development of \sys{}, a new GPU-cluster simulator that is network-contention-aware using GPU-free characterization and contention-aware network model.
\item Production trace-driven evaluation showing that \sys{} substantially reduces JCT simulation error compared to existing simulators.%
\end{itemize}

\section{Background}

\subsection{DT Job}\label{sec:2.1}

A DT job uses multiple GPU workers to train a model in parallel, reducing training time and scaling to larger models or datasets \cite{shin2026prediction}. In data parallel training~\cite{wang2024towards}, each worker holds a model replica and processes a disjoint partition of each mini-batch. After the backward pass, workers synchronize gradients through all-reduce so that they apply identical parameter updates.
Each iteration consists of 1) \emph{compute time} $T^{\mathrm{comp}}_J$ for the forward and backward passes on local GPUs and 2) \emph{networking time $T^{\mathrm{net}}_J$} for gradient exchange; their sum is the \emph{iteration time} $T^{\mathrm{iter}}_J = T^{\mathrm{comp}}_J + T^{\mathrm{net}}_J$. Since a job completes $W_J$ iterations, its total training time is $T^{\mathrm{train}}_J = W_J \cdot T^{\mathrm{iter}}_J$. 

The key metric for evaluating a DT job is JCT, which includes both waiting time, defined as the time from job submission until GPU allocation, and training time. Makespan is another key metric for evaluating scheduling quality, which is defined as the time from the first job submission to the completion of the last job in a workload.

\subsection{DT Job Scheduling and Contention}\label{sec:2.2}
A job scheduler (e.g., Kubernetes~\cite{kubescheduler} and Volcano~\cite{volcano2025}) allocates GPUs to submitted jobs and decides how their workers are scheduled across GPU servers. This scheduling decision determines which server NICs and inter-server links carry each job's network traffic. When concurrent jobs share the same NICs or inter-server links, their traffic can contend for shared bandwidth; we refer to these jobs as co-running jobs.

\begin{figure}[t]
    \centering
    \includegraphics[width=.8\linewidth]{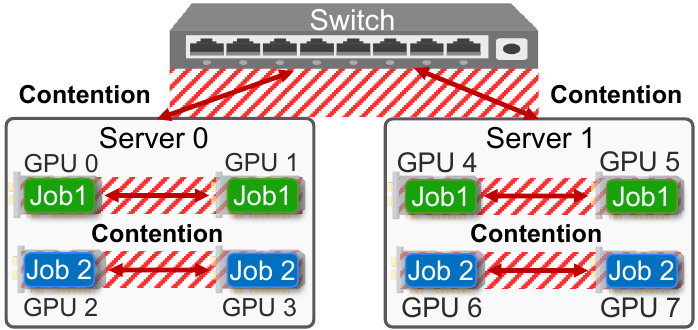}
    \caption{Network contention example.}
    \label{fig:1}\vspace{-1em}
\end{figure}

Fig.~\ref{fig:1} illustrates how scheduling creates such network contention. The green job occupies GPUs 0 and 1 on Server 0, and GPUs 4 and 5 on Server 1, and the blue job occupies GPUs 2 and 3 on Server 0, and GPUs 6 and 7 on Server 1. Each job generates networking traffic to synchronize gradients across GPUs. 
So, intra-server flows (e.g., between GPUs 0–1 and GPUs 2–3) and inter-server flows (e.g., between GPUs 0–4 and GPUs 2–6) overlap, thereby causing network contention at both the intra- and inter-server levels. Thus, scheduling determines not only which GPUs a job uses, but also which network resources its all-reduce traffic shares with other jobs.

There exist diverse scheduling policies. For example, Kubernetes, de facto job orchestration framework, provides two representative policies \cite{kubescheduler}. Bin packing (or MostAllocated in Kubernetes) schedules a job on the fewest servers possible, preferring compact allocation but spanning multiple servers when necessary. Conversely, Load balancing (or LeastAllocated in Kubernetes) spreads a job's workers across as many servers as possible to balance GPU usage.

\subsection{Existing GPU Cluster Simulators} \label{subsec:existing-sim}

Running controlled scheduling experiments on real GPU clusters is expensive and hard to reproduce, so researchers often rely on trace-driven simulation to compare scheduling policies~\cite{tiresias,pollux,gavel,muri}. A cluster-level simulator replays a production trace~\cite{philly,acme} under different scheduling policies and compares the resulting JCT and makespan. Each trace entry specifies a job's submission time and GPU demand, but such traces typically do not contain the training time of each target workload. The simulator therefore assigns each job a workload from a curated model pool, such as common vision and language models, and fills in its training time by conducting profiling on jobs, which measure each workload running alone on a fixed number of GPUs. The simulator advances through discrete events, including job arrivals, GPU allocations, and job completions. At each allocation event, the scheduling policy decides which queued jobs receive GPUs and how their workers are scheduled across GPU servers \cite{shin2026prediction}.

Existing simulators fall into two categories. \emph{No-contention simulators} use each job's standalone training time without adjustment, implicitly assuming that co-running jobs do not affect one another. Tiresias~\cite{tiresias} replays fixed job durations from recorded traces without modeling contention. Gavel~\cite{gavel} profiles each job across heterogeneous GPU types for resource allocation, but does not model network contention of jobs.

In contrast, \emph{static-contention simulators} approximate network contention by a fixed factor. Pollux~\cite{pollux} and Muri~\cite{muri}, for example, increase training time by 10\% to account for contention. However, the same ratio is applied regardless of a job's networking volume, other co-running jobs, or the server NICs and inter-server links shared by synchronization traffic. Thus, existing simulators do not capture the workload- and scheduling-dependent nature of network contention.

\section{Motivation}
\label{sec:motivation}

This section presents our motivating experiments. We first describe the experiment setup and then analyze the results.

\subsection{Experiment Setup}
\label{sec:mot:env}

\subsubsection{Baselines}
We evaluate Tiresias~\cite{tiresias} and Pollux~\cite{pollux} as representative no-contention and static-contention simulators, respectively. Tiresias uses each job's standalone training time regardless of co-running jobs. Pollux applies a fixed contention ratio 10\% to co-running jobs. Other simulators, such as Gavel~\cite{gavel} and Muri~\cite{muri}, follow similar designs and show trends consistent with Tiresias and Pollux, respectively; we omit their results due to space constraints.

\subsubsection{Metrics}
We evaluate the fidelity of simulation results by comparing them against ground-truth measurements. We use mean absolute percentage error (MAPE) across four metrics: average training time, average JCT, 99-th percentile (P99) JCT, and makespan. The metrics are explained in \S\ref{sec:2.1}.

\begin{table}[t]
    \centering
    \caption{Models used in trace.}
    \label{tab:models}
    \footnotesize
    \setlength{\tabcolsep}{3pt}
    \renewcommand{\arraystretch}{0.95}
    \begin{tabular}{@{}p{0.22\columnwidth}p{0.55\columnwidth}
                    >{\raggedleft\arraybackslash}p{0.15\columnwidth}@{}}
        \toprule
        Dataset & Model(s) & Batch size \\
        \midrule
        SQuAD~\cite{squad}   & BERT-base~\cite{bert}  & 4--32 \\
        iMDB~\cite{imdb}     & GPT-2~\cite{gpt2}      & 4--32 \\
        Synthetic            & Whisper~\cite{whisper}  & 4--32 \\
        \midrule
        CIFAR-10~\cite{cifar10}
            & \makecell[l]{AlexNet~\cite{alexnet},
                DenseNet-40/100~\cite{densenet},\\
                ResNet-44/110~\cite{resnet}}
            & \makecell[r]{128--\\32768} \\
        \midrule
        ImageNet~\cite{imagenet}
            & \makecell[l]{Inception-v3~\cite{inceptionv3},
                ResNet-50~\cite{resnet},\\
                VGG-16~\cite{vgg16}, GoogLeNet~\cite{googlenet}}
            & \makecell[r]{64--\\2048} \\
        \bottomrule
    \end{tabular}\vspace{-1em}
\end{table}

\subsubsection{Workloads}
We prepare a 6.6-hour DT job trace sampled from the Microsoft Philly trace~\cite{philly}. Since the public trace does not provide model details, we use 12 representative models spanning computer vision, natural language processing, and speech domains, as summarized in Table~\ref{tab:models}. The trace contains 30 jobs requesting 2--8 GPUs, with per-job durations ranging from 3 to 424 minutes. All jobs use data-parallel training with ring all-reduce for gradient synchronization. Job submission times follow a Poisson process with arrival rate $\lambda=0.005$, which makes GPUs nearly fully utilized.
We use Kubernetes load balancing, which spreads jobs across nodes with more available resources (\S\ref{sec:2.2}). Other policies show similar trends.

\subsubsection{Machine}
We run the DT job trace on the two existing simulators and compare their outputs against ground-truth measurements from a physical GPU cluster. The simulators run on a separate host equipped with an Intel Xeon E5-2650 v4 @ 2.20~GHz CPU, 24 cores, and 64~GB of DDR4 memory; this host is sufficient for simulation and does not become a bottleneck. To obtain ground-truth measurements for MAPE, we run the same trace on a physical testbed provided by Lambda Cloud \cite{lambdalabs}. The testbed consists of two 8-GPU V100 servers, each equipped with eight V100-SXM2-16GB GPUs, two Intel Xeon Gold 5220R CPUs, and 440 GiB of DRAM. The two servers are connected through a 5 Gbps interface.

\subsection{Simulation Fidelity}

\label{sec:motivation-error}

Table \ref{tab:motivation_existing_error} shows the simulation errors of existing simulators against the physical testbed. Tiresias incurs large errors across all metrics: 58.71\% for average training time, 73.64\% for average JCT, 67.19\% for P99 JCT, and 47.51\% for makespan. These errors arise because Tiresias ignores the training time increase caused by co-running jobs. Once training time is incorrectly modeled, the error propagates to GPU release times, queued-job waiting times, and eventually JCT and makespan.

Pollux reduces the errors only modestly by applying a fixed contention ratio. However, its MAPE remains high: 56.92\% for average training time, 72.23\% for average JCT, 65.31\% for P99 JCT, and 46.42\% for makespan. This shows that a static contention factor is insufficient to capture network contention. Errors of this magnitude make simulation results unreliable for scheduler evaluation. Next, we show that network contention is the root cause of the significant errors.

\begin{table}[t]
\centering
\caption{Fidelity of existing simulators: MAPE (\%, $\downarrow$: better).}
\label{tab:motivation_existing_error}
\setlength{\tabcolsep}{3pt}
\begin{tabular}{lcc}
\toprule
\multicolumn{1}{c}{}
 & \makecell{Tiresias (No-contention)} & \makecell{Pollux (Static-contention)}\\
\midrule
Average training time & 58.71 & 56.92 \\
Average JCT           & 73.64 & 72.23 \\
P99 JCT               & 67.19 & 65.31 \\
Makespan              & 47.51 & 46.42 \\
\bottomrule
\end{tabular}
\vspace{-1em}
\end{table}

\subsection{Cause of Poor Fidelity}
\label{sec:motivation-time}

\begin{figure}[t]
    \centering
    \begin{subfigure}[b]{0.4\columnwidth}
        \includegraphics[width=\columnwidth]{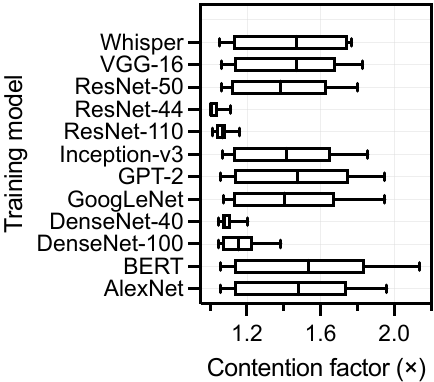}
        \caption{Iteration-time changes.}\label{fig:mot2:a}
    \end{subfigure}
    \hfil
    \begin{subfigure}[b]{0.58\columnwidth}
        \includegraphics[width=\columnwidth]{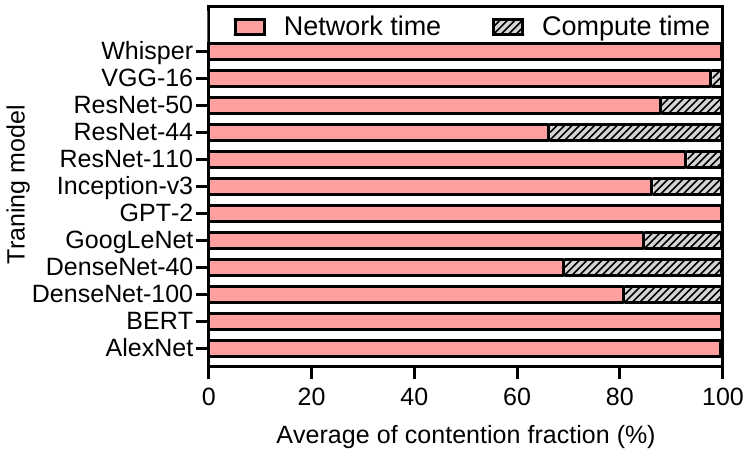}
        \caption{Breakdown of changes.}\label{fig:mot2:b}
    \end{subfigure}
    \caption{Root-cause analysis of poor fidelity.}
    \label{fig:motivation}\vspace{-1.5em}
\end{figure}

To identify why existing simulators present significant errors, we run the following experiments on the physical testbed using the same 12-workload pool. Because training time and JCT accumulate over repeated DT iterations (\S\ref{sec:2.1}), we analyze how per-iteration time changes by contention.

Each experiment runs two 8-GPU jobs concurrently across two servers with fixed worker locations. On each server, the measured job $A$ uses four GPUs, and the co-running job $B$ disjointly uses the other four GPUs. Both jobs perform gradient synchronization across the two servers, so their inter-server networking shares the same NICs. For each job $A$, we change the co-running job $B$ among the 12 models in Table \ref{tab:models} and measure the per-iteration time of $A$.

Then, we define the iteration-time change of job $A$ when co-running with job $B$ as $T_{\mathrm{pair}}(A, B)/T_{\mathrm{solo}}(A)$,
where $T_{\mathrm{pair}}(A, B)$ is the mean iteration time of job $A$ while co-running with $B$, and $T_{\mathrm{solo}}(A)$ is the mean iteration time of $A$ when running alone. A value of 1 indicates no change, whereas values $>$ 1 indicate that the iteration time of $A$ increases.

Fig. \ref{fig:mot2:a} shows the iteration-time change of each measured job as the co-running job varies across the 12 models. Across the models, the iteration-time changes are 1.33$\times$ on average and reach up to 2.14$\times$ (BERT model). 
Fig. \ref{fig:mot2:b} decomposes the iteration-time change into the part caused by network contention and compute contention. On average, network time accounts for 88.9\% of the change, while compute accounts for only 11.1\%. Even for the model with the largest compute fraction, networking still explains 66.4\% of the change.

The results show that the network and its contention are the dominant sources of changes in iteration time. As per-iteration time errors accumulate into job-level timing metrics such as training time and JCT, modeling network contention is key for improving simulation fidelity.

\section{Design}
\label{sec:design}

\begin{figure}[t]
    \centering
    \includegraphics[width=\linewidth]{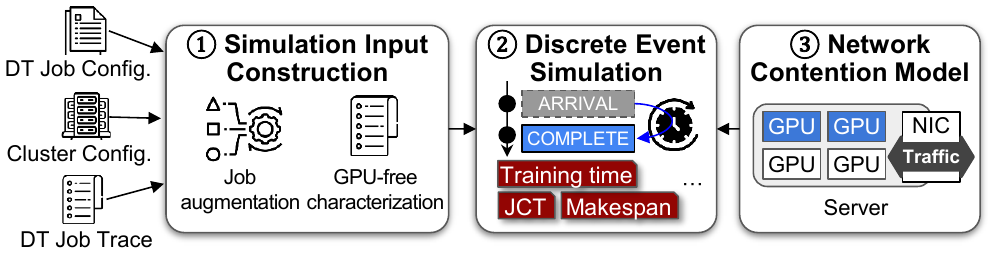}\vspace{-.7em}
    \caption{Overview of \sys{}.}
    \label{fig:overview}\vspace{-1em}
\end{figure}

Fig.~\ref{fig:overview} shows the workflow of \sys{}. \sys{} takes DT job configurations, cluster configurations, and a job trace as input (\S\ref{subsec:input}). First, \sys{} runs simulation input construction (① in Fig.~\ref{fig:overview}), which prepares all required information to conduct DT job simulation. GPU cluster simulators take public traces in general; however, typical public traces do not include each job's details, such as the model type, dataset, per-iteration compute time, network time, or networking volume. Simulation input construction of \sys{} fills the missing information from pre-defined job pools and GPU-free characterization that gets job-level details without requiring any GPUs (details in \S\ref{subsec:job-level-sim}).

Next, the discrete event simulation (② in Fig.~\ref{fig:overview}) replays the constructed input over time. It processes job arrivals and completions, applies the scheduling policy of the cluster, updates job progress, and schedules each job's estimated completion event (\S\ref{subsec:des}). Depending on the scheduling, the simulation invokes network contention model (explained next) to update each affected job's iteration time. After running all jobs in the input, the simulator outputs the training time and JCT for each job, as well as the overall makespan.

The network contention model (③ in Fig.~\ref{fig:overview}) captures network contention and estimates its impact on iteration time. Given the current worker assignment on GPUs and servers, which the scheduler determines at simulator run time, the model estimates how contention from co-running jobs inflates each job's network time and thus its iteration time (\S\ref{subsec:cluster-level-sim}).

\subsection{Simulation Input Construction}\label{subsec:input}
\sys{} takes three inputs: DT job trace, DT job configuration, and cluster configuration.
\sys{} runs simulations from a DT job trace. Representative sources of such traces include public production traces from large-scale GPU clusters \cite{philly,helios,acme}. We denote the trace as a set of DT jobs, $\mathcal{T}=\{J\}$, where each job initially carries only the metadata provided by these traces: $J=\langle a_J,\, g_J,\, d_J\rangle$, including its submission time $a_J$, the number of requested GPUs $g_J$, and its trace-reported duration $d_J$. Production traces rarely include detailed training workload information essential for simulation, such as 1) workload description $m_J$ of the deep learning model type, dataset, training framework, and collective communication algorithm, 2) per-iteration compute time $T^{\mathrm{comp}}_J$, 3) networking time $T^{\mathrm{net}}_J$, and 4) networking volume $V^{\mathrm{net}}_J$.

To make the traces usable for simulation, \sys{} performs two steps. First, it augments each trace job with the missing $m_J$, following the common practice in other simulators \cite{tiresias,gavel,pollux,muri}. Second, it characterizes the job to obtain the remaining details, $T^{\mathrm{comp}}_J$, $T^{\mathrm{net}}_J$, and $V^{\mathrm{net}}_J$. The first augmentation is achieved using another input, job configuration, which we explain next. The second job characterization is explained in \S\ref{subsec:job-level-sim}.

\subsubsection{Job Augmentation}
To augment the details, we take DT job configuration as another input which provides the candidate workload descriptions used for the first augmentation step. Since production traces omit $m_J$, \sys{} takes candidate DT job configurations as separate inputs and uses them as a pool from which the missing $m_J$ is sampled. Each candidate configuration is provided as a YAML file and specifies the job details that form $m_J$, including the deep learning model type, dataset, training framework, and collective communication algorithm. If users do not provide custom configurations, \sys{} uses its default DT job configurations, which consist of the models in Table \ref{tab:models}. For each trace job, \sys{} samples an $m_J$ from this pool and attaches it to the job, yielding $J=\langle a_J,\, g_J,\, d_J,\, m_J\rangle$.

Another input, cluster configuration, describes the target cluster specifications and scheduling policy used in the simulation. It includes the number of servers, the number of GPUs per server, the per-server network bandwidth, and the scheduling policy. \sys{} supports heterogeneous cluster configurations, allowing users to specify different GPU types, numbers of GPUs per server, and per-server network bandwidths. \sys{} also provides implementations of several scheduling policies and allows users to implement custom policies, which are used in the simulation (\S\ref{subsec:des}).

\subsubsection{GPU-Free Characterization}
\label{subsec:job-level-sim}
In addition to job augmentation, replaying a DT job trace requires per-job execution characteristics that public traces omit (\S\ref{subsec:input}) and cannot be filled by augmentation. Existing simulators recover these by profiling each job for a few iterations on a separate GPU cluster, measuring its iteration time when the job runs alone \cite{pollux}.\footnote{Since iteration time depends on the job details, simulators profile it after augmenting each trace job with a DT job configuration.} This approach is costly and hardware-dependent: it occupies real GPUs and must be repeated whenever the GPU type, model pool, or training configuration changes. \sys{} avoids this cost through \emph{GPU-free characterization}, which replaces real-GPU profiling with a job-level simulator.

Job-level simulators have been developed in the GPU architecture domain to model single-job execution. For example, ASTRA-sim \cite{astra-sim} and SimAI \cite{simai} simulate a DT job at operator-level granularity across different GPU architectures. Assuming the job runs alone on the requested number of GPUs ($g_J$), they estimate the mean compute time and networking time, which together constitute the iteration time, and expose operator-level network transfer records from which the networking volume can be derived. \sys{} uses such a simulator to generate a single-job profile for each trace job, which then serves as input to the cluster-level simulation.

We use ASTRA-sim\footnote{\sys{} is not tied to ASTRA-sim; it can use any single-job simulator that reports per-iteration compute time, networking time, and networking volume.} as the job-level backend, as it is known to be accurate in the compute time, networking time, and networking volume of DT jobs.
For each job $J$, \sys{} obtains its per-iteration compute time $T^{\mathrm{comp}}_J$, networking time $T^{\mathrm{net}}_J$, and networking volume $V^{\mathrm{net}}_J$. 

\sys{} keeps compute time and networking time separate because it applies network contention only to the networking part of an iteration. This is consistent with our observation in \S\ref{sec:motivation-time}: iteration time changes almost due to networking time (and its contention), while compute time remains unchanged.

The networking volume $V^{\mathrm{net}}_J$ is the total network traffic generated by all workers of job $J$ during one iteration. Rather than measuring it on real hardware, \sys{} extracts it from the operator-level transfer records, summing the transmitted and received bytes reported by ASTRA-sim's networking operators across all workers.

\sys{} also determines each job's iteration count $W_J$. Treating the trace-reported duration $d_J$ as the standalone runtime, \sys{} sets $W_J = d_J / T^{\mathrm{iter}}_J$. With $W_J$ replacing $d_J$, each job is now fully described as:
\begin{equation}
J=\langle a_J,\, g_J,\, m_J,\, W_J,\, T^{\mathrm{comp}}_J,\, T^{\mathrm{net}}_J,\, V^{\mathrm{net}}_J\rangle.
\label{eq:job-tuple}
\end{equation}

The constructed input (including job description) is further used to estimate how network contention changes each job's iteration time, as explained in \S\ref{subsec:cluster-level-sim}.

\subsection{Discrete Event Simulation}
\label{subsec:des}
\sys{} replays the constructed input as a discrete-event simulation that processes, in time order, two events per job: an \textsc{Arrive} event at its submission time $a_J$, and a \textsc{Complete} event scheduled at its estimated finish time $\hat{f}_J$ once the job is scheduled. A job scheduler applies the scheduling policy from the cluster configuration (\S\ref{subsec:input}) to decide which waiting jobs run and where their workers are assigned. \sys{} updates the cluster state only when either of these two event types occurs, as these are the only moments when GPU allocation or NIC sharing can change.

When a job arrives at $a_J$, the scheduler assigns it to free GPUs if it fits; otherwise the job joins a waiting queue. When a job completes, it releases its GPUs, and the scheduler then revisits the waiting queue to assign any job that fits. Either way, when a job is assigned or completed at simulation time $t$, the set of jobs sharing each NIC can change. Thus, \sys{} computes the contention factor $S_J(t)$ of every affected job using the network contention model (to be explained in \S\ref{subsec:cluster-level-sim}).

Based on the contention factor from the networking contention model, \sys{} models the iteration time of job $J$ at simulation time $t$ as:
\begin{equation}
\tau_J(t) = T^{\mathrm{comp}}_J + T^{\mathrm{net}}_J \cdot S_J(t).
\label{eq:iter}
\end{equation}

Only the networking term is scaled by $S_J(t)$ because network contention affects networking time but not compute time (\S\ref{sec:motivation-time}).

At each event time $t_{now}$, \sys{} first advances every active job by the iterations it has completed since its previous update time $t_{last}$. Since the set of running jobs is fixed between two consecutive events, $S_J(t)$ and hence $\tau_J(t)$ remain constant over that interval. The completed iterations are:
\begin{equation}
\Delta N_J = \frac{t_{now} - t_{last}}{\tau_J(t_{last})}.
\end{equation}

\sys{} then decrements the remaining iterations:
\begin{equation}
W_J(t_{now}) = \max\!\left(0,\ W_J(t_{last}) - \Delta N_J\right).
\end{equation}

After processing the event and applying the scheduling policy, \sys{} recomputes $S_J(t_{now})$ for affected jobs and projects their new finish times as:
\begin{equation}
\hat{f}_J = t_{now} + W_J(t_{now})\,\tau_J(t_{now}).
\end{equation}
The corresponding \textsc{Complete} event is scheduled or updated to $\hat{f}_J$. Advancing time event by event, rather than through fixed time ticks, lets \sys{} track contention that changes over each job's lifetime.

\textbf{Simulation output.}
Once the last \textsc{Complete} event has been processed, \sys{} reports per-job and cluster-level metrics. For each job, training time is the time spent executing after GPU allocation, while JCT is the time from job submission to completion, including both the waiting time before GPU allocation and the training time after allocation. At the cluster level, makespan is the time from the first job's submission to the last job's completion across the whole trace.

\subsection{Networking Contention Model}
\label{subsec:cluster-level-sim}
The discrete event simulation in the previous subsection uses the contention factor $S_J(t)$ to update each job's iteration time. This subsection explains how \sys{} computes this factor from the current worker assignment and per-job networking profile. The network contention model does not determine the worker assignment. Instead, the assignment is produced by the job scheduler at run time during the discrete-event simulation (\S\ref{subsec:des}). This subsection assumes that the current assignment is given and defines how the contention factor is computed.

We denote job $J$'s worker assignment---the assignment of its $N_W=g_J$ workers to servers---by $\pi_J$, and the number of $J$'s workers on server $i$ by $n_W^i$. Using the per-job profile $(T^{\mathrm{comp}}_J, T^{\mathrm{net}}_J, V^{\mathrm{net}}_J)$ from \S\ref{subsec:job-level-sim} together with $\pi_J$, \sys{} estimates how much each job's iteration time changes when its synchronization traffic shares server NICs and inter-server links with co-running jobs. \sys{} captures this as a per-job \emph{contention factor}, derived from each job's bandwidth demand on every server NIC as follows.

\subsubsection{Bandwidth Demand}
\sys{} converts each job's networking volume and scheduling decision into a bandwidth demand on each server NIC, which is used to determine the contention factor. We assume synchronous data-parallel training with ring all-reduce.\footnote{This assumption only affects bandwidth-demand in Eq.~\ref{eq:tex}. \sys{} can support other parallelism strategies by replacing this equation with a strategy-specific communication-demand, without changing the rest of the simulator.} In each iteration, every worker of job $J$ sends and receives the same amount of gradient data, so for a job with networking volume $V^{\mathrm{net}}_J$ spread over $N_W$ workers, the per-worker bandwidth demand is:
\begin{equation}
c = V^{\mathrm{net}}_J / N_{W}.
\label{eq:perlink}
\end{equation}

In ring all-reduce, the workers form a logical ring, and each worker communicates only with its two neighbors in the ring. Two neighboring workers on the same server exchange data locally and do not use the NIC, whereas two neighboring workers on different servers communicate via the NIC connecting the servers. NCCL orders the ring so that workers on a given server are placed next to each other~\cite{nccl}; as a result, the workers on server $i$ form a continuous block of the ring, and only the two workers at the ends of that block communicate with workers on other servers. Each of these two boundary links carries the per-worker demand $c$, so a job whose workers span multiple servers places a demand of $2c$ on server $i$'s NIC. The one exception is a two-worker job ($N_W = 2$) split across two servers: the ring then reduces to a single link between the two workers, so the demand is $c$ instead of $2c$. Combining these, the inter-server bandwidth demand of job $J$ on server $i$ is:
\begin{equation}
T_{ex}^i = \begin{cases}
    \min(1,\, N_W - n_W^i) \times c \times 2 & N_W > 2 \\
    \min(1,\, N_W - n_W^i) \times c           & N_W = 2
\end{cases}
\label{eq:tex}
\end{equation}
where $n_W^i$ is the number of $J$'s workers on server $i$. The quantity $N_W - n_W^i$ counts $J$'s workers assigned to other servers, so $\min(1, N_W - n_W^i)$ is $1$ when at least one such worker exists, meaning $J$'s traffic crosses server $i$'s boundary, and $0$ when all of $J$'s workers are on server $i$. In the latter case $T_{ex}^i = 0$ and $J$ uses no inter-server NIC bandwidth.

\subsubsection{Contention Factor}
\sys{} then computes how much bandwidth each job actually receives at each server NIC, treating the NICs independently. Let $D_J^i = T_{ex}^i$ be job $J$'s demand on server $i$, $C$ the NIC capacity, and $\mathcal{J}^i(t)$ the set of active jobs sending traffic through that NIC at time $t$. If the combined demand of these jobs fits within $C$, each job receives its full demand; if the combined demand exceeds $C$, the NIC bandwidth is split among them in proportion to their demands. Both cases are captured by:
\begin{equation}
A_J^i(t) =
D_J^i \cdot
\min\!\left(1,\ \frac{C}{\sum_{k \in \mathcal{J}^i(t)} D_k^i}\right),
\label{eq:alloc}
\end{equation}
where the $\min(1, \cdot)$ term equals $1$ when the NIC is not oversubscribed and scales every job's allocation down by the same ratio when it is. The contention factor of $J$ at server $i$ then measures how far its allocation falls short of its demand:
\begin{equation}
S_J^i(t) = \frac{D_J^i}{\max(A_J^i(t),\ \texttt{MIN\_BW})},
\label{eq:s_i}
\end{equation}
where \texttt{MIN\_BW} is a small lower bound on the allocation that keeps the ratio finite. The factor equals $1$ when $J$ receives all the bandwidth it needs, and rises above $1$ as contention reduces $J$'s allocation, lengthening $J$'s networking time by the same proportion. Because synchronous all-reduce advances only as fast as the slowest group of workers, a job's overall contention factor is the largest factor across the servers it occupies:
\begin{equation}
S_J(t) = \textstyle\max_i S_J^i(t).
\label{eq:s_job}
\end{equation}
A job with no inter-server traffic does not suffer contention, so $S_J(t) = 1$.

\section{Evaluation}

\begin{table*}[t]
\centering
\caption{Simulation fidelity: MAPE (\%) across training time, JCT metrics, and makespan (\S\ref{sec:eval-accuracy-aggregate}, $\downarrow$: better).}
\label{tab:trace1-mape}
\small
\setlength{\tabcolsep}{4pt}
\begin{tabular}{l|rrrrr|rrrrr}
\toprule
 & \multicolumn{5}{c|}{Bin packing} & \multicolumn{5}{c}{Load balancing} \\
\cmidrule(lr){2-6}\cmidrule(lr){7-11}
Method & \shortstack{Average\\training time} & \shortstack{Average\\ JCT} & \shortstack{Median\\JCT} & \shortstack{P99\\JCT} & Makespan & \shortstack{Average\\training time} & \shortstack{Average\\JCT} & \shortstack{Median\\JCT} & \shortstack{P99\\JCT} & Makespan \\
\midrule
Tiresias & 38.19 & 36.38 & 34.98 & 50.42 & 53.45 & 56.75 & 53.93 & 54.20 & 65.28 & 65.44 \\
Pollux & 35.30 & 33.00 & 31.83 & 48.46 & 51.08 & 54.72 & 51.42 & 51.86 & 63.89 & 63.64 \\
\rowcolor{gray!15} \sys{} & \textbf{19.74} & \textbf{11.10} & \textbf{1.62} & \textbf{24.84} & \textbf{17.61} & \textbf{17.05} & \textbf{21.83} & \textbf{19.60} & \textbf{8.38} & \textbf{7.72} \\
\bottomrule
\end{tabular}
\end{table*}

We implement \sys{} in Rust and Python (4.4K lines of source code) and use ASTRA-sim \cite{astra-sim} for GPU-free characterization. Based on the implementation, we first describe the experiment setup (\S\ref{sec:eval:setup}) and then present the results. Our implementation is publicly available\footnote{https://github.com/OSSS-KU/MoSim}.

\subsection{Experiment Setup} \label{sec:eval:setup}
We extend the setup in \S\ref{sec:mot:env} to a larger scale. We reuse the same model pool and evaluation methodology, but scale the workload from one 6.6-hour trace of 30 jobs to a 13-hour trace of 60 jobs, derived from the Philly trace \cite{philly}. We also scale the testbed from two to four 8-GPU V100 servers (32 GPUs in total), keeping the same per-server hardware and inter-server bandwidth as in \S\ref{sec:mot:env}. As in \S\ref{sec:mot:env}, we use measurements on this physical testbed as the ground truth.

\noindent\textbf{Baselines.}
We compare \sys{} against the same two simulators as \S\ref{sec:mot:env}: Tiresias \cite{tiresias}, no-contention simulator that replays each job's standalone time, and Pollux \cite{pollux}, static-contention simulator that inflates every co-running job's time by 10\%.

\noindent\textbf{Scheduling policies.}
We evaluate \sys{} under two cluster scheduling policies: bin packing, which places workers on the fewest servers possible by preferring servers with the fewest free GPUs, and load balancing, which spreads workers across servers by preferring servers with the most free GPUs. Both are provided by Kubernetes \cite{kubescheduler}.

We evaluate the following items:
\begin{itemize}[leftmargin=1em]
\item \textbf{Simulation fidelity (\S\ref{sec:eval-accuracy}):} 
how closely \sys{} reproduces training time, JCT, and makespan measured on the physical testbed. We report 1) aggregate fidelity using MAPE for these metrics;
and 2) distribution-level fidelity using Kolmogorov--Smirnov and Wasserstein distances between the simulated and measured, ground-truth JCT distributions.%
\item \textbf{Fidelity of network contention model (\S\ref{sec:eval-design}):} compare the contention factor from \sys{} against the ground-truth value measured on the physical testbed in \S\ref{sec:motivation-time}.%
\item \textbf{Input construction overhead (\S\ref{sec:eval-cost}):} the resource overhead of constructing each simulator's inputs for the models. We report the required machine time, monetary cost, and normalized cost, comparing real-GPU profiling in the baselines with GPU-free characterization in \sys{}.
\end{itemize}

\begin{table}[t]
\centering
\caption{Simulation fidelity: distribution-level fidelity, KS and Wasserstein distance of JCT distribution  (\S\ref{sec:eval-accuracy-distance}, $\downarrow$: better).}
\label{tab:dist-metrics-combined}
\small
\setlength{\tabcolsep}{2.5pt}
\begin{tabular}{l|rr|rr}
\toprule
 & \multicolumn{2}{c|}{Bin packing} & \multicolumn{2}{c}{Load balancing} \\
\cmidrule(lr){2-3}\cmidrule(lr){4-5}
Method & KS distance & Wasserstein (s) & KS distance & Wasserstein (s)\\
\midrule
Tiresias & 0.32 & 3499 & 0.38 & 7082 \\
Pollux & 0.27 & 3184 & 0.37 & 6753 \\
\rowcolor{gray!15} \sys{} & \textbf{0.13} & \textbf{1579} & \textbf{0.22} & \textbf{2864} \\
\bottomrule
\end{tabular}
\end{table}

\subsection{Simulation Fidelity}\label{sec:eval-accuracy}
\subsubsection{Aggregate Fidelity}\label{sec:eval-accuracy-aggregate}

Table~\ref{tab:trace1-mape} shows the MAPE of training time, JCT (average, median, and P99), and makespan for Tiresias, Pollux, and \sys{} under both bin packing and load balancing scheduling policies. \sys{} achieves the best (lowest) MAPE in all cases, showing the highest accuracy across all metrics and scheduling policies.
For training time, \sys{} improves accuracy by 1.93$\times$ over Tiresias and 1.79$\times$ over Pollux under bin packing, and by 3.33$\times$ and 3.21$\times$ respectively under load balancing.

For average JCT, \sys{} improves accuracy by 3.28$\times$ over Tiresias and 2.97$\times$ over Pollux under bin packing. Under load balancing, \sys{} achieves 2.47$\times$ and 2.36$\times$ improvements, respectively. For median JCT, \sys{} achieves improvements of 21.59$\times$ over Tiresias under bin packing and 2.77$\times$ under load balancing. For P99 JCT, \sys{} improves accuracy by 2.03$\times$ over Tiresias and 1.95$\times$ over Pollux under bin packing. Under load balancing, \sys{} achieves larger improvements of 7.79$\times$ and 7.62$\times$ respectively.

For makespan, \sys{} improves accuracy by 3.04$\times$ over Tiresias and 2.9$\times$ over Pollux under bin packing, and by 8.48$\times$ and 8.25$\times$ respectively under load balancing.

\subsubsection{Distribution-Level Fidelity}\label{sec:eval-accuracy-distance}
Here, we evaluate how accurately \sys{} captures the overall JCT distribution of the trace. We use two distance metrics between the measured and simulated JCT distributions.
First, we measure Kolmogorov--Smirnov (KS) distance, which quantifies the maximum vertical distance between the measured and simulated cumulative distribution functions (CDFs) as $\max_x |F_{\text{measured}}(x) - F_{\text{simulated}}(x)|$. This metric captures the largest distributional gap at any point and is useful for detecting bias across quantiles, including the tail region.
Second, we measure Wasserstein distance, which quantifies how far the simulated JCT distribution is from the measured JCT distribution in terms of magnitude. 
Unlike KS distance, which captures the largest CDF gap, Wasserstein distance reflects overall JCT differences across the distribution.

Table \ref{tab:dist-metrics-combined} shows that \sys{} consistently outperforms the baselines across both distance metrics. Under bin packing, \sys{} achieves a KS distance of 0.13, which is 2.46$\times$ better than Tiresias and 2.08$\times$ better than Pollux. It also achieves a Wasserstein distance of 1579 s, improving over Tiresias by 2.22$\times$ and Pollux by 2.02$\times$. These results indicate that the simulated distribution closely follows the measured distribution across all quantiles. Under load balancing, \sys{} obtains a KS distance of 0.22, which is 1.73$\times$ better than Tiresias and 1.68$\times$ better than Pollux, and a Wasserstein distance of 2864 s, which improves over Tiresias by 2.47$\times$ and Pollux by 2.36$\times$. The results show that \sys{} provides superior distribution-level fidelity, accurately capturing not only aggregate statistics but also the overall shape and spread of the JCT distribution.

\subsection{Fidelity of Network Contention Model}

\label{sec:eval-design}

\begin{figure}[t]
    \centering
    \begin{subfigure}[b]{0.58\columnwidth}
        \centering
        \includegraphics[width=\linewidth]{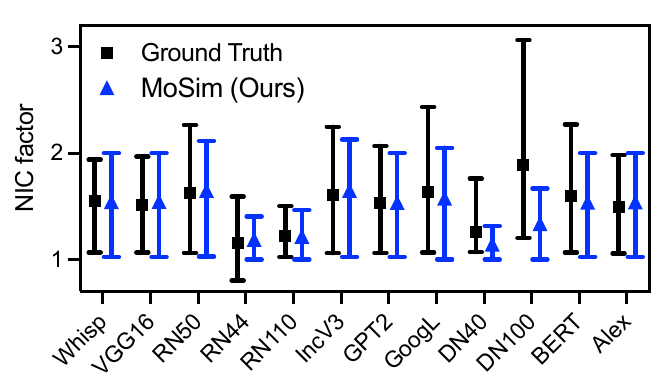}
        \caption{NIC contention factor comparison.}\label{fig:eval-ncm-a}
    \end{subfigure}
    \hfil
    \begin{subfigure}[b]{0.3\columnwidth}
        \centering
        \small
        \setlength{\tabcolsep}{3pt}
        \begin{tabular}{l|r}
        \toprule
        Method & MAPE (\%)\\
        \midrule
        Tiresias & 29.69 \\
        Pollux & 21.05 \\
        \rowcolor{gray!15} \sys & \textbf{8.63} \\
        \bottomrule
        \end{tabular}
        \vspace*{1mm}
        \vspace{6mm}
        \caption{MAPE(\%).}\label{fig:eval-ncm-b}
    \end{subfigure}
    \caption{Fidelity of network contention model: NIC contention factor comparision and MAPE (\S\ref{sec:eval-design}, MAPE $\downarrow$: better).}
    \label{fig:eval-nic-contention-factor}\vspace{-.3em}
\end{figure}

We next evaluate the fidelity of \sys{}'s network contention model. We reuse the pairwise job-running measurements from \S\ref{sec:motivation-time}, where one job is fixed, and its co-running job varies across the 12 models. From the measurements, we calculate the real contention factor and compare it with the estimate from the network contention model.

Fig.~\ref{fig:eval-nic-contention-factor} shows the results. In Fig.~\ref{fig:eval-ncm-a}, for each model on the x-axis, we show two whiskers: one black whisker for the ground-truth value and one blue whisker for the value estimated by our network contention model. For each model, there are 12 counterpart models, so the whiskers show the range of all values. We observe that the estimates from \sys{} are highly similar to the ground-truth values. This corresponds to the MAPE of 8.63\% for \sys{} across 144 possible experiments, as shown in Fig.~\ref{fig:eval-ncm-b}. We also calculate the MAPE values of Tiresias and Pollux, which set the fixed contention factor to 1.0 and 1.1, respectively. Compared with Tiresias and Pollux, \sys{}'s MAPE is 3.44$\times$ and 2.44$\times$ lower, respectively.

\begin{table}[t]
\centering
\caption{Input construction overhead: time (hours) on GPU-machine and CPU-machine, cost (\$), and normalized cost (\S\ref{sec:eval-cost}, $\downarrow$: better).}
\label{tab:cost}
\small
\setlength{\tabcolsep}{3pt}
\begin{tabular}{lrrrr}
\toprule
Method & \makecell{GPU-hours} & \makecell{CPU-hours}
         & Cost (\$) & \makecell{Normalized cost} \\
\midrule
Tiresias & 41.2 & -
    & 32.55 & 44.6$\times$ \\
Pollux & 41.2 & -
    & 32.55 & 44.6$\times$ \\
\rowcolor{gray!15}
\sys{} & - & \textbf{2.7}
    & \textbf{0.73} & \textbf{1}$\times$ \\
\bottomrule
\end{tabular}
\vspace{-.5em}
\end{table}

\subsection{Input Construction Overhead} \label{sec:eval-cost}
Table~\ref{tab:cost} compares the overhead of each simulator, especially for constructing simulation inputs. We use on-demand cloud prices for the analysis: \$6.32/h for the GPU machine, i.e., an 8-GPU V100 VM~\cite{lambdalabs}, and \$0.2688/h for the CPU machine, i.e., a t4g.2xlarge CPU instance in AWS us-east-2.
Existing simulators, such as Tiresias and Pollux, run profiling to fill in missing job details, consuming the GPU machine for about 41 hours. Instead, \sys{} runs its GPU-free characterization on the CPU machine for about 2.7 hours. As a result, \sys{} reduces the monetary cost of input construction by 44.6$\times$.

\section{Related Work}

\textbf{GPU cluster simulators.}
Existing GPU cluster simulators differ from \sys{} along two criteria: how they model network contention and how they obtain per-job characteristics. As discussed in \S\ref{subsec:existing-sim}, prior simulators either ignore contention (Tiresias \cite{tiresias}, Gavel \cite{gavel}) or apply a fixed penalty ratio (Pollux \cite{pollux}, Muri \cite{muri}), and all obtain per-job characteristics through real-GPU profiling. \sys{} differs on both: it models contention dynamically from placement and networking volume, and obtains the same per-job characteristics through GPU-free characterization, removing GPU dependence from simulation.

\textbf{Single-job simulators.}
\sys{} relies on a single-job simulator as the backend for its GPU-free characterization (\S\ref{subsec:job-level-sim}). ASTRA-sim \cite{astra-sim}, SimAI \cite{simai}, and Multiverse \cite{multiverse} simulate a single job at operator-level granularity and are complementary to \sys{}, which operates at the cluster level. \sys{} uses one such simulator (ASTRA-sim) to replace per-job profiling and layers its cluster-level contention model on top; any backend that reports per-iteration compute time, networking time, and networking volume can serve in its place.

\textbf{Communication schedulers.}
A separate line of work mitigates network contention directly. \textsc{Cassini} \cite{cassini} reduces network contention through fine-grained delay scheduling, and Crux \cite{crux} assigns flow priorities by GPU intensity. VALO accurately and efficiently splits datacenter traffic of AI workloads across multiple paths \cite{valo}. These techniques intervene to reduce contention, whereas \sys{} models and predicts it; the two are independent, and such mitigation could itself be included and modeled within \sys{}.

\section{Conclusion}
This paper introduces \sys{}, a GPU-cluster simulator that models DT jobs under dynamic network contention. \sys{} combines GPU-free characterization with network contention model, obtaining each job's compute time, networking time, and networking volume without real-GPU profiling, and modeling how shared server NICs and inter-server links change each job's iteration time with scheduling decisions. Compared with existing simulators, \sys{} reduces simulation error by up to 3.28$\times$ for average JCT, 7.79$\times$ for P99 JCT, and 8.48$\times$ for makespan. It also estimates NIC contention factors with 8.63\% MAPE and reduces input construction cost by 44.6$\times$. The results show that network contention modeling is important for faithful GPU-cluster simulation.

\section*{Acknowledgment}
Yeonho Yoo, Hyunho Lee, and Hyunmok Choi contributed equally. This research was supported by National Research Foundation of Korea (NRF) grant funded by Korea government (MSIT) (RS-2024-00336564), by Institute of Information \& Communications Technology Planning \& Evaluation (IITP) grant funded by Ministry of Science and ICT (MSIT) (RS-2026-25518394), by ICT Creative Consilience Program through IITP grant funded by MSIT (IITP-2026-RS-2020-II201819), by IITP under the Artificial Intelligence Convergence Innovation Human Resources Development grant funded by Korea government (MSIT) (IITP-2026-RS-2023-00254592), by ANCHOR through Seoul ANCHOR Center funded by MOE and Seoul Metropolitan Government (2026-ANCHOR-01-003-09), and by computing support from Lambda Cloud. Corresponding authors: Gyeongsik Yang and Chuck Yoo.

\bibliographystyle{IEEEtran}
\bibliography{ref}

\end{document}